\documentclass[numberedappendix]{emulateapj}

\defcitealias{1959flme.book.....L}{Landau-Lifshitz}

\DeclareMathAlphabet{\emf}{U}{euf}{m}{n}
\DeclareMathAlphabet{\ebf}{U}{euf}{b}{n}
\DeclareMathAlphabet{\emn}{U}{eur}{m}{n}
\DeclareMathAlphabet{\ebn}{U}{eur}{b}{n}
\DeclareMathAlphabet{\ems}{U}{eus}{m}{n}
\DeclareMathAlphabet{\ebs}{U}{eus}{b}{n}

\newcommand{\kB}{k_\mathrm{B}}

\newcommand{\me}{m_\mathrm{e}}
\newcommand{\dT}{\delta_\mathrm{T}}
\newcommand{\gs}{g_\mathrm{s}}
\newcommand{\Rr}{\ems{R}_\mathrm{r}}
\newcommand{\ar}{a_\mathrm{r}}
\newcommand{\Gf}{\ems{G}_\mathrm{f}}
\renewcommand{\bf}{b_\mathrm{f}}

\begin{document}

\title{A Class of Physically Motivated Closures for Radiation Hydrodynamics}

\shorttitle{Physical Radiation Hydrodynamic Closures}

\author{Chi-kwan Chan}

\affil{Institute for Theory and Computation,
  Harvard-Smithsonian Center for Astrophysics,
  60 Garden Street, Cambridge, MA 02138} 

\email{ckchan@cfa.harvard.edu}

\begin{abstract}
  Radiative transfer and radiation hydrodynamics use the relativistic
  Boltzmann equation to describe the kinetics of photons.  It is
  difficult to solve the six-dimensional time-dependent transfer
  equation unless the problem is highly symmetric or in equilibrium.
  When the radiation field is smooth, it is natural to take angular
  moments of the transfer equation to reduce the degrees of freedom.
  However, low order moment equations contain terms that depend on
  higher order moments.  To close the system of moment equations,
  approximations are made to truncate this hierarchy.  Popular
  closures used in astrophysics include flux limited diffusion and the
  $M_1$ closure, which are rather \textit{ad hoc} and do not
  necessarily capture the correct physics.  In this paper, we propose
  a new class of closures for radiative transfer and radiation
  hydrodynamics.  We start from a different perspective and highlight
  the consistency of a fully relativistic formalism.  We present a
  generic framework to approximate radiative transfer based on
  relativistic \citeauthor{Grad1949}'s moment method.  We then derive
  a 14-field method that minimizes unphysical photon self-interaction.
\end{abstract}


\section{Introduction}

Radiative transfer and radiation hydrodynamics use the
ultra-relativistic Boltzmann transport equation to describe the
kinetics of photon.  The radiative intensity, which is proportional to
the photon distribution function, is a seven-dimensional hypersurface
embedded in eight-dimensional phase space \citep{1973grav.book.....M}.
Unless the radiative intensity is in equilibrium, or the problem is
highly symmetric, it is difficult to solve the radiative transfer
equation either analytically or numerically \citep[see standard
  textbook such as][]{1960ratr.book.....C, 1984oup..book.....M,
  1986rpa..book.....R, 1991par..book.....S, 2001irtm.book.....P,
  2004rahy.book.....C, 2006cmt..conf.....G}.  On the one hand, there
is no successful theory to reduce the complexity of radiative
transfer.  On the other hand, although numerical algorithms such as
direct discretization of the radiative transfer equation and Monte
Carlo methods exist, they are computationally too expensive.

In astrophysics, very often we are interested in the radiative energy
and flux instead of the intensity.  If the radiation field is smooth
(in terms of directions), we simply take angular moments of the
radiative transfer equation and solve only for the frequency-dependent
moment equations.  The $\mbox{zeroth-}$, $\mbox{first-}$, and second
order angular moments of the intensity carry clear physical meanings.
They are the radiative energy density, radiative flux, and radiative
stress tensor, respectively.  For many classic problems in
astrophysics like stellar atmospheres, the global symmetry of the
system is used to further reduce the degrees of freedom.  Analytical
or numerical solutions are then obtained by solving the reduced
frequency-dependent moment equations.  This approach is naturally
extended to radiation hydrodynamics, which not only describes how the
(moving) media radiates, but also how radiation feeds back to matter
\citep{1984oup..book.....M, 2001JQSRT..71...61M, 2004rahy.book.....C,
  2007ApJ...667..626K}.

The dynamic equations of the radiative stress tensor contain terms
that are related to the third order moment of the intensity, while the
equations of the third order moments depend on the fourth order
moments and so on.  In order to close the system, we need to make
approximations and truncate the moment hierarchy.  This is known as
the \emph{closure problem} in radiative transfer.  Popular closure
schemes in astrophysics include variant forms of flux limited
diffusion \citep[FLD,][]{1970JCoPh...6....1C, 1981ApJ...248..321L,
  1983JQSRT..29..223P, 1984JQSRT..31..149L}, $P_N$ approximations
\citep[sometimes with diffusion corrections, see][and reference
  therein]{2000JQSRT..64..619O, 2009CMT....21..511S,
  2008JCoPh.227.2864M, 2010JQSRT.111.2052R}, the $M_1$ closure
\citep[equivalent to maximal entropy closure,
  see][]{1992A&A...265..345J, 2000A&A...356..559S,
  2008PhRvD..78b4023F}, and variable Eddington factors
\citep[VEF,][]{1964SAOSR.167..108F, 1969JQSRT...9..407P,
  1970MNRAS.149...65A, 2008PASJ...60..377F, 2008PASJ...60.1209F,
  2009PASJ...61..367F}\footnote{The term VEF has recently been abused
  in the literature.  The original \citet{1970MNRAS.149...65A}
  algorithm uses an iterative solver for the Eddington factors, which
  is technically not a closure relation.  The closures such as $M_1$
  should really be called moment methods with \emph{prescribed
    Eddington factors} \citep{1987ApJ...323..227F,
    1990ApJ...354...83P, 1992A&A...266..613K}.}.  They have been
successfully applied to many astrophysics problems, yet their
underlying assumptions (what is the optimal form of a flux limiter for
a particular problem?  why is the entropy of photons maximized?  how
many moments do we need?) are rather \textit{ad hoc} and do not
guarantee to capture the correct physics.

In hydrodynamics, the Chapman-Enskog theory expands the particle
distribution in power series of the Knudsen number, which is the ratio
between the particle mean free path and the typical length scale in
the problem.  The zeroth order expansion gives rise to ideal
hydrodynamics, while the first order terms allow us to calculate
transport coefficients such as kinematic viscosity and thermal
conductivity \citep{1965ApJ...141..201H, 1970mtnu.book.....C,
  1979ittk.book.....L, 2002rbet.book.....C}.  We would like to apply a
similar technique to radiative transfer in order to solve the closure
problem.  Unfortunately, photons do not self-interact.  The photon
mean free path is determined by the extinction coefficient, which is a
material property.  In the cases when the photon density is high but
the media is optically thin, radiation dominates the energy budget but
cannot be thermalized.  The Chapman-Enskog series for photon
distribution function fails to converge \citep{2001JQSRT..71...61M}.

In addition to the closure problem, as studies of astrophysical
objects become more detailed, complicated sub-structures such as
convection in stars and turbulence in accretion disks are always
found.  The radiating medium moves non-uniformly, which makes the
computation of spectra more complicated due to the spatial-dependent
Doppler effect.  The standard approach is to Taylor expand the
extinction and emission coefficients and to derive moment equations to
at least $\mathcal{O}(v/c)$, where $v$ and $c$ are the speed of the
media and the speed of light.  The resulting equations have terms that
are physically important even in the limit $v/c \rightarrow 0$
\citep{1984oup..book.....M}.  Keeping track of these higher order
terms in different physical regimes is a challenging task
\citep{2007ApJ...667..626K}.  Moreover, there are astrophysical
systems such as the inner regions of accretion disks and
ultra-relativistic jets that move with relativistic speed so the
described approach converges too slowly.

If radiative feedback is weak, we can solve the hydrodynamic equations
and compute the radiative spectra using post-processing
\citep[see][]{2007CQGra..24..259N, 2009ApJ...701..521C}.  However,
there is a large set of problems for which the feedback is important.
We have to solve the frequency-dependent three-dimensional moment
equations at every single time step.  The full system of radiation
hydrodynamics is highly non-linear.  Studies of turbulence show that
reducing the number of spatial dimensions and enforcing symmetry in
these systems can produce fundamentally wrong results
\citep{1967PhFl...10.1417K, 1971JFM....47..525K}.

To summarize, we have listed three major difficulties\footnote{Another
  fundamental difficulty of radiation hydrodynamics comes from the
  large separation between the radiative time scale and the (hydro-)
  dynamic time scale.  Nevertheless, this difficulty raises not
  because of the moment method.  It is a generic property of
  non-relativistic astrophysical systems.  We will therefore leave it
  out from this paper.} in using moment methods for radiative transfer
and radiation hydrodynamics:
\begin{enumerate}
  \item Because of the collisionless nature of photons, standard
    methods in kinetic theory such as the Chapman-Enskog theory fails
    to provide a physical closure.
  \item When the non-uniform motion of media is considered, it is
    difficult to take properly the spatial-dependent Doppler effect
    into account.
  \item If the radiative feedback is important, the naive approach is
    computationally too expensive to couple radiation back to
    hydrodynamics.
\end{enumerate}

Hydrodynamics depends only on the frequency-integrated radiative
force.  Many existing codes, therefore, reduce the degrees of freedom
and resolve problem~(iii) by using frequency-integrated equations
\citep{1992ApJS...80..819S, 2001ApJS..135...95T, 2003ApJS..147..197H,
  2007A&A...464..429G, 2007ApJ...667..626K, 2008MNRAS.387..295A,
  2008PhRvD..78b4023F}.  If the radiative spectrum is needed, it is
always possible to post-process the numerical solution.

Once we integrate over frequency, the moment equations are fully
relativistic.  The radiative energy density, momentum, and stress
tensors together form a covariant stress-energy tensor.  We can then
ignore the complication of mixed frame formalism.  This relativistic
approach is originally developed to solve radiative transfer in
general relativity \citep{1972ApJ...171..127A, 1978Ap&SS..56..191S,
  1981MNRAS.194..439T, 1982MNRAS.199.1137U}.  It is well adopted in
numerical general relativistic hydrodynamics
\citep{2005gr.qc.....3085A, 2006MNRAS.367.1739P, 2007MNRAS.382.1041T,
  2008PhRvD..78b4023F}.  However, we believe that its true advantage
is in the covariant equations, which trivially addresses problem~(ii).

\citet{Grad1949} proposed a moment method which expands the ratio
between the particle distribution and the equilibrium distribution in
a multi-dimensional polynomials of the momentum.  Instead, it treats
the moments as fundamental fields and keeps track of their evolutions.
This method is independent of the mean free path.  Once the reference
frame and temperature are chosen, \citeauthor{Grad1949}'s coefficients
are simply linear transforms of the momentum moments, which converge
exponentially fast when the distribution function is well
behaved\footnote{It is a simple application of Darboux's principle.
  See, for example, \citet{Boyd1999}, on convergent properties of
  series expansions.}.  Because photon transport is linear,
\citeauthor{Grad1949}'s moments method provides the most natural way
to resolve problem~(i).

The idea of applying \citeauthor{Grad1949}'s moment method to study
radiation is not new.  A classic paper by \citet{1981MNRAS.194..439T}
presented very detailed moment formalisms for relativistic radiative
transfer by using projected, symmetric, trace-free tensors
\citep{1980RvMP...52..299T}.  Shortly after that,
\citet{1982MNRAS.199.1137U} derived a 14-field approximation for
general relativistic radiative transfer based on
\citeauthor{Grad1949}'s method.  However, as far as we know, these
attempts were only used in one-dimensional problems
\citep[e.g.][]{1996MNRAS.281.1183Z}.  On the one hand, this is
possibly due to the general concept that Newtonian formalisms are
always simpler than relativistic formalisms.  On the other hand,
\citeauthor{1981MNRAS.194..439T} and
\citeauthor{1982MNRAS.199.1137U}'s 14-field methods are indeed
complicated compared to the standard $\mathcal{O}(v/c)$ equations in
\citet{1984oup..book.....M}.  Their advantages only appear when we
consider more physical regimes such as the ones described in
\citet{2007ApJ...667..626K}.

Based on the above points and some additional physical understanding
of moment methods, we propose a new class of closure schemes for
radiative transfer and radiation hydrodynamics in this paper.  The
paper is organized as the following.  In the next section, we
introduce our notations and review the standard radiation
hydrodynamics equations.  In section~\ref{sec:Grad}, we describe
\citeauthor{Grad1949}'s moment method and its ultra-relativistic
generalization.  We also summarize the linear transformations that we
derive in the appendixes.  In section~\ref{sec:physics}, we study
carefully ideal hydrodynamics and some existing closures for radiative
transfer.  They provide us important physical insight, which leads to
a new class of closure schemes.  Although the scheme is generic, we
specifically look at the 14-field approximations and derive the
closure equations in section~\ref{sec:14}.  Finally, we conclude this
paper in section~\ref{sec:conclusions}.

\section{Notations and Standard Equations}
\label{sec:notations}

We use the component notation in \citet{1973grav.book.....M} though
out this paper.  Greek indices run from 0 to 3 and Latin indices run
only from 1 to 3.  The metric tensor is denoted by $g^{\alpha\beta}$
and the metric signature is $(-,+,+,+)$.  A point in spacetime is
denoted by $x^\alpha = (c\,t, x^i)$ with $c$ being the speed of light.
The Einstein summation convention is used unless specified otherwise.

We start from the standard radiative transfer equation.  Let $\nu$ be
the photon frequency, $n^i$ be a three-dimensional unit vector.  We
define $I_\nu \equiv I(t, x^i; \nu, n^i)$ be the specific radiative
intensity, $\eta_\nu$ be the total emission coefficient, and
$\chi_\nu$ be the total extinction coefficient.  The intensity is then
governed by the radiative transfer equation
\begin{equation}
  \left(\frac{1}{c}\frac{\partial}{\partial t} +
        n^i \frac{\partial}{\partial x^i}\right) I_\nu =
  \eta_\nu - \chi_\nu I_\nu.
  \label{eq:transfer}
\end{equation}

We first describe the standard moment methods
\citep{1960ratr.book.....C, 1978stat.book.....M, 1984oup..book.....M,
  1986rpa..book.....R, 1991par..book.....S, 2001irtm.book.....P,
  2004rahy.book.....C}.  Taking the zeroth and first angular momentum
of the whole transfer equation~(\ref{eq:transfer}), we obtain two
frequency-dependent moment equations,
\begin{eqnarray}
  \partial_t E_\nu +       \partial_i F_\nu^i    & = & - c   G^0_\nu
    \label{eq:1st}\\
  \partial_t F_\nu^i + c^2 \partial_j P_\nu^{ij} & = & - c^2 G^i_\nu
    \label{eq:2nd}
\end{eqnarray}
The quantities on the left are the frequency-dependent radiative
energy density, radiative flux, and radiative stress tensor,
\begin{eqnarray}
  E_\nu      & \equiv & \frac{1}{c} \int\!d\Omega\ I_\nu, \\
  F_\nu^i    & \equiv &             \int\!d\Omega\ I_\nu n^i, \\
  P_\nu^{ij} & \equiv & \frac{1}{c} \int\!d\Omega\ I_\nu n^i n^j.
\end{eqnarray}
The ones on the right are the radiative energy and moment inputs (to
the radiating medium),
\begin{eqnarray}
  G^0_\nu & \equiv & \frac{1}{c}
    \int\!d\Omega\left(\chi_\nu I_\nu - \eta_\nu\right), \\
  G^i_\nu & \equiv & \frac{1}{c}
    \int\!d\Omega\left(\chi_\nu I_\nu - \eta_\nu\right) n^i.
\end{eqnarray}

Note that the four-tuple $(E_\nu, F_\nu^i)$ is not a four-vector so
equations~(\ref{eq:1st}) and (\ref{eq:2nd}) do not form a covariant
equation.  The quantities $E_\nu$, $F_\nu^i$, and $P_\nu^{ij}$, as
well as $G_\nu^0$ and $G_\nu^i$, all depend on the reference frame.
This is exactly the difficulty in solving angular moment equations
when the radiating medium moves non-uniformly --- we cannot Lorentz
transform $(E_\nu, F_\nu^i)$ to obtain their values in a different
frame.

We follow \citet{1984oup..book.....M} to derive the covariant formulas
as we suggested in the introduction.  Let $h$ be Planck's constant and
$n^\alpha \equiv (1, n^i)$ be the ``unit'' null vector.  The photon
four-momentum is $p^\alpha \equiv (h\nu/c)n^\alpha$.  The Lorentz
invariant photon distribution function $f$ is related to the intensity
by the equation
\begin{equation}
  f(x^\alpha, p^\beta) \equiv f(t, x; \nu, n^i) =
  \frac{c^2}{h^4 \nu^3}\ I(t, x; \nu, n^i).
\end{equation}
We can now rewrite equation~(\ref{eq:transfer}) as a
ultra-relativistic Boltzmann transport equation,
\begin{equation}
  p^\alpha f_{;\alpha} = \emn{e} - \emn{x} f,
  \label{eq:Boltzmann}
\end{equation}
where the subscript $_{;\alpha}$ denotes covariant derivative.  The
Lorentz invariant emission coefficient
\begin{equation}
  \emn{e}(\nu) \equiv \frac{c}{h^3\nu^2} \eta_\nu
\end{equation}
and extinction coefficient
\begin{equation}
  \emn{x}(\nu) \equiv \frac{h\nu}{c} \chi_\nu
\end{equation}
are usually assumed independent of $f$.

Using the covariant volume element, $d^3p / p^0$, we can take
four-momentum moments of equation~(\ref{eq:Boltzmann}) to arbitrary
order,
\begin{equation}
  {R^{\alpha_1 \alpha_2 \cdots \alpha_{l+1}}}_{\!;\alpha_{l+1}} = 
  - G^{\alpha_1 \alpha_2 \cdots \alpha_l}. \label{eq:rad}
\end{equation}
The moment tensor on the left hand side is defined by
\begin{equation}
  R^{\alpha_1 \alpha_2 \cdots \alpha_{l+1}} \equiv c \int\!\frac{d^3p}{p^0}
  f\,p^{\alpha_1} p^{\alpha_2} \cdots p^{\alpha_{l+1}},
\end{equation}
while on the right hand side, we have
\begin{equation}
  G^{\alpha_1 \alpha_2 \cdots \alpha_l} \equiv c \int\!\frac{d^3p}{p^0}
  \,(\emn{x} f - \emn{e}) p^{\alpha_1} p^{\alpha_2} \cdots p^{\alpha_l}.
\end{equation}
We refer $G^{\alpha_1 \alpha_2 \cdots \alpha_l}$ as the \emph{moment
  extinction term} because $-G^{\alpha_1 \alpha_2 \cdots \alpha_l}$ is
usually called the \emph{moment production term}
\citep{2002rbet.book.....C}.

It is easy to connect the relativistic formulas with the
non-relativistic ones.  Using the first order equation, the radiative
stress-energy four-tensor is
\begin{equation}
  R^{\alpha\beta} \equiv
  \frac{1}{c} \int\!d\nu\,d\Omega\, I_\nu n^\alpha n^\beta =
  \left[\begin{array}{cc}
    E       & F^{j}/c \\
    F^{i}/c & P^{ij}
  \end{array}\right].
  \label{eq:Rab}
\end{equation}
The frame-dependent components are the frequency-integrated radiative
energy density, radiative flux, and radiative stress tensor.
Similarly, the radiative four-force is
\begin{equation}
  G^\alpha \equiv \frac{1}{c} \int\!d\nu\,d\Omega\,(\chi_\nu I_\nu -
  \eta_\nu) n^\alpha = \left[\begin{array}{c}G^0
    \\ G^i\end{array}\right].
\end{equation}
The temporal and spatial components are the frequency-integrated
radiative energy and momentum inputs.  We remark that higher angular
moments of the intensity, even though they are frequency-integrated,
do not have counterparts in equation~(\ref{eq:rad}).

For completeness, we also write down the equations for relativistic
hydrodynamics.  We use $v^i$ to denote the fluid velocity.  The
four-velocity is $u^\alpha = \gamma (c, v^i)$, where $\gamma$ is the
Lorentz factor.  The material four-momentum is given by
\begin{equation}
  T^\alpha = \rho u^\alpha,
\end{equation}
where $\rho$ is material density in the local Lorentz rest frame.  For
simplicity, we assume perfect fluid so that the material stress-energy
tensor takes the simple form
\begin{equation}
  T^{\alpha\beta} = \rho h u^\alpha u^\beta + p g^{\alpha\beta}.
  \label{eq:stress-energy}
\end{equation}
Here, $h = e/c^2 + p/\rho c^2$ is the specific enthalpy, $e$ is the
specific internal energy density including the rest energy, and $p$ is
the thermal pressure.

After all the definitions, standard radiation hydrodynamics can be
summarized in three tensor equations, namely, the continuity equation,
\begin{equation}
  {T^\alpha}_{\!;\alpha} = 0,
  \label{eq:cont}
\end{equation}
the covariant Euler equation,
\begin{equation}
  {T^{\,\alpha\beta}}_{\!;\beta} = G^\alpha,
  \label{eq:Euler}
\end{equation}
and the radiative stress-energy equation
\begin{equation}
  {R^{\alpha\beta}}_{\!;\beta} = - G^\alpha.
  \label{eq:rad1}
\end{equation}
The above equations are Lorentz covariant.  Nevertheless, we refer
them as ``lab frame equations'' in order to distinguish from the
radiation fiducial frame and the fluid comoving frame.

\section{Grad's Moment Method}
\label{sec:Grad}

Because of the perfect fluid assumption, the material part of standard
radiation hydrodynamics is closed by equation~(\ref{eq:stress-energy})
and an equation of state.  We will refer this as the ideal
(relativistic) hydrodynamic closure.  Conversely, there are nine
unknowns in the trace-free symmetric radiative stress-energy tensor
but only four independent equations.  The transport part of radiation
is subject to the closure problem.  In addition, it is unclear how to
relate the radiative four-force to other macroscopic quantities.
\citet{Grad1949} proposed evolving the moments as fundamental fields
and use them to reconstruct the particle distribution function.  The
unknown components of the moments and the moment extinction terms can
then be evaluated self-consistently.  In this section, we will
generalize \citeauthor{Grad1949}'s moment method to work with
radiation.

\subsection{Grad's Expansion}

We generalize \citeauthor{Grad1949}'s moment method in a covariant
form and expand $f$ as the follow multi-dimensional power series
\begin{equation}
  f = f^{(0)} \left(
    \hat{a} + \hat{a}_\alpha p^\alpha +
    \hat{a}_{\alpha\beta} p^\alpha p^\beta + \cdots \right)
    \label{eq:Grad_dimensional}
\end{equation}
at each point $x^\alpha$.  In the above expansion, we choose
\begin{equation}
  f^{(0)} \equiv \frac{2 / h^3}{\exp(-p^\alpha U_\alpha / \theta) - 1}
  \label{eq:blackbody}
\end{equation}
be the equilibrium (black body) photon distribution function, where
$U^\alpha$ is a four-velocity of some fiducial observer and $\theta$
is the energy scale for the equilibrium distribution.  The
coefficients $\hat{a}$, $\hat{a}_\alpha$, $\hat{a}_{\alpha\beta}$,
... are totally symmetric with trace-free spatial parts (see
appendix~\ref{app:Grad} or \citealt{1980RvMP...52..299T}).

In the fiducial reference frame, $- p^\alpha U_\alpha$ reduces to the
photon energy $h \nu$.  Using the energy scale $\theta$, we define a
dimensionless quantity
\begin{equation}
  \xi \equiv h\nu/\theta.
\end{equation}
The equilibrium distribution reduces to $f^{(0)} = (2/h^3)w(\xi)$,
where
\begin{equation}
  w(\xi) \equiv \frac{1}{\exp(\xi) - 1}
  \label{eq:weight}
\end{equation}
can be interpreted as a dimensionless weight.  We can rewrite
\citeauthor{Grad1949}'s expansion as
\begin{equation}
  f = \frac{2}{h^3}\, w(\xi)
    \Bigl( a + \xi a_\alpha n^\alpha +
    \xi^2 a_{\alpha\beta} n^\alpha n^\beta + \cdots \Bigr).
    \label{eq:Grad}
\end{equation}
The rescaled coefficients $a$, $a_\alpha$, $a_{\alpha\beta}$, ... are
all dimensionless.

Note that the weight $w(\xi)$ depends explicitly on direction after a
Lorentz transformation --- this is simply the Doppler effect.  The
coefficients in expansions~(\ref{eq:Grad_dimensional}) and
(\ref{eq:Grad}) are specific for the (at the moment, arbitrarily)
chosen energy scale and fiducial frame.  Nevertheless, the infinite
series contains full information of the original Lorentz invariant
distribution $f$.  We can Taylor expand the anisotropic part of the
weight and obtain another set of \citeauthor{Grad1949}'s coefficients
in the other frame.  The distribution is fiducial frame dependent only
after we truncate the expansion at certain order.

\subsection{Solving for Grad's Coefficients}
\label{sec:coef}

In principle, we could derive evolution equations for
\citeauthor{Grad1949}'s coefficients by substituting
\citeauthor{Grad1949}'s expansion~(\ref{eq:Grad}) into the
ultra-relativistic Boltzmann equation~(\ref{eq:Boltzmann}) but we
would then lose the physical intuition.  Instead, we follow
\citet{Grad1949} and treat the moments as fundamental fields.  The
left hand sides of the moment equations~(\ref{eq:rad}) suggests that
we need to keep track of $R^{\alpha_1\alpha_2\cdots\alpha_l0}$
(components that have at least one 0 in the indices).  The flux terms
$R^{i_1 i_2 \cdots i_{l+1}}$ (components with no 0 in the indices) and
the extinction terms $G^{\alpha_1\alpha_2\cdots\alpha_l}$ must be
solved in terms of them.

Taking moment is a linear operation.  We can skip the distribution
function $f$ and directly write down a linear transformation between
the coefficients and the moments.  Let the Euler script
\begin{equation}
  \ems{R}^{\alpha_1\alpha_2\cdots\alpha_l} \equiv
  \frac{\lambda^3}{c}\biggl(\frac{c}{\theta}\biggr)^{l-1}
  R^{\alpha_1\alpha_2\cdots\alpha_l}
  \label{eq:rescale}
\end{equation}
be the dimensionless moments, where $\lambda \equiv hc/\theta$ is the
photon mean separation.  Using appendix~\ref{app:moments}, the
relation between the dynamic variables and \citeauthor{Grad1949}'s
coefficients can be summarized in the following hierarchy of matrix
equations.
\begin{equation}
  \left[\begin{array}{l}
    \ems{R}^0\\\ems{R}^{00}\\\ems{R}^{000}\\\ \vdots
  \end{array}\right]
  = 8\pi
  \left[\begin{array}{cccc}
    W_2 & W_3 & W_4 & \cdots\\
    W_3 & W_4 & W_5 & \cdots\\
    W_4 & W_5 & W_6 & \cdots\\
    \vdots & \vdots & \vdots & \ddots
  \end{array}\right]
  \left[\begin{array}{l}a\\a_0\\a_{00}\\\ \vdots\end{array}\right],
  \label{eq:R0a0}
\end{equation}
\begin{equation}
  \left[\begin{array}{l}
    \ems{R}^{0i}\\\ems{R}^{00i}\\\ \vdots
  \end{array}\right]
  = \frac{8\pi}{3}
  \left[\begin{array}{ccc}
    W_4 & 2 W_5 & \cdots\\
    W_5 & 2 W_6 & \cdots\\
    \vdots & \vdots & \ddots
  \end{array}\right]
  \left[\begin{array}{l}a_i\\a_{0i}\\\ \vdots
  \end{array}\right],
  \label{eq:Riai}
\end{equation}
\begin{equation}
  \left[\begin{array}{l}\ems{R}^{0ij}\\\ \vdots\end{array}\right]
  \!\!-\!\frac{1}{3}\!
  \left[\begin{array}{l}\ems{R}^{000}\\\ \vdots\end{array}\right]
  \!\!\dT^{ij}\! = \!\frac{16\pi}{15}\!\!
  \left[\begin{array}{cc}
    W_6 & \cdots\\
    \vdots & \ddots
  \end{array}\right]
  \left[\begin{array}{l}a_{ij}\\\ \vdots\end{array}\right],
  \label{eq:Rijaij}
\end{equation}
where $W_l$ is the shorthand of the integral $W_l \equiv \int_0^\infty
d\xi w(\xi) \xi^l$.  Substituting our weighting
function~(\ref{eq:weight}), the integral has the closed from solution
$W_l \equiv \Gamma(l+1)\zeta(l+1)$.  Some numerical values are listed
in equation~(\ref{eq:Wl_values}).

Of course, there are infinitely many equations in the hierarchy.  We
just present the ones that are useful for this paper.  It is easy to
solve the coefficients by inverting the above transforms.  We will
provide some examples in section~\ref{sec:physics} and \ref{sec:14}.

\subsection{Computing the Flux Terms}
\label{sec:flux}

Once we obtain \citeauthor{Grad1949}'s coefficients, we can use them
to compute the flux terms.  \citeauthor{Grad1949}'s moment method is
linear in the fiducial frame.  The linearity naturally form a class of
closure schemes.  Since we fix the weight $w(\xi)$, the only freedoms
in the closures are the energy scale $\theta$ and the fiducial
reference frame corresponds to $U^\alpha$.

We will discuss how to choose the fiducial velocity $U^\alpha$ in
section~\ref{sec:physics}.  For now, we use the subscript
$_\mathrm{r}$ to indicate the radiation fiducial frame,
\begin{equation}
  R_\mathrm{r}^{\alpha_1\alpha_2\cdots\alpha_{l}} =
    {\Lambda_\mathrm{r}^{\alpha_1}}_{\!\beta_1}
    {\Lambda_\mathrm{r}^{\alpha_2}}_{\!\beta_2} \cdots
    {\Lambda_\mathrm{r}^{\alpha_l}}_{\!\beta_l}
    R^{\beta_1\beta_2\cdots\beta_l},
\end{equation}
where ${\Lambda_\mathrm{r}^\alpha}_\beta$ is the Lorentz
transformation matrix.  We use the result derived in
appendix~\ref{app:moments} to compute the flux terms.
\begin{equation}
  \Rr^i
  = \frac{8\pi}{3}\Bigl(W_3 {\ar}_i + 2 W_4 {\ar}_{0i} 
  + 3 W_5 {\ar}_{00i} + \dots\Bigr),
\end{equation}
\begin{equation}
  \Rr^{ij}
  = \frac{16\pi}{15}\Bigl(W_5 {\ar}_{ij} + 3 W_6 {\ar}_{0ij}
  + \dots\Bigr) + \frac{1}{3}\Rr^{00}\dT^{ij},
\end{equation}
\begin{equation}
  \Rr^{ijk}
  = \frac{16\pi}{35}\Bigl(W_7 {\ar}_{ijk} + \dots\Bigr)
  + \frac{1}{5}\Bigl(\Rr^{00i}\dT^{jk}
  + \Rr^{00j}\dT^{ki} + \Rr^{00k}\dT^{ij}\Bigr).
\end{equation}
To obtain the flux terms in the lab frame, we simply apply an inverse
Lorentz transformation to
$R_\mathrm{r}^{\alpha_1\alpha_2\cdots\alpha_l}$.

Note that the dimensionless photon number density $\Rr^0$ does not
enter the closure equations, we can use it to choose the energy scale.
We require $\theta$ to approach $\kB T$ in local thermal dynamic
equilibrium, i.e., $\Rr^0 = 8\pi W_2$, which leads to
\begin{equation}
  \theta = \left(\frac{h^3 c^2 R_\mathrm{r}^0}{8\pi W_2}\right)^{1/3}.
  \label{eq:theta}
\end{equation}

\subsection{Computing the Extinction Terms}
\label{sec:extinction}

Photon emission and absorption are fluid properties.  They take their
simplest form in the fluid comoving frame.  Therefore, we use the
fluid temperature $\kB T$ and the fluid four-velocity $u^\alpha$ to
replace $\theta$ and $U^\alpha$ in \citeauthor{Grad1949}'s expansion
(see appendix~\ref{app:extinction}).

To compute the extinction terms, we apply Lorentz transformation to
boost the moments to the fluid comoving frame,
\begin{equation}
  G_\mathrm{f}^{\alpha_1\alpha_2\cdots\alpha_{l}} =
    {\Lambda_\mathrm{f}^{\alpha_1}}_{\!\beta_1}
    {\Lambda_\mathrm{f}^{\alpha_2}}_{\!\beta_2} \cdots
    {\Lambda_\mathrm{f}^{\alpha_l}}_{\!\beta_l}
    G^{\beta_1\beta_2\cdots\beta_l}.
\end{equation}
We follow the same procedure described in section~\ref{sec:coef} to
obtain the comoving coefficients $a_\mathrm{f}$,
${a_\mathrm{f}}_{\alpha}$, ${a_\mathrm{f}}_{\alpha\beta}$, ...  The
subscript $_\mathrm{f}$ here indicates the fluid comoving frame.
Introducing the notation
\begin{equation}
  {\bf}_{\alpha_1\alpha_2\cdots\alpha_l} =
    {a_\mathrm{f}}_{\alpha_1\alpha_2\cdots\alpha_l} - \delta^{l0}
\end{equation}
and letting the Euler script
\begin{equation}
  \Gf^{\alpha_1\alpha_2\cdots\alpha_l} \equiv
  \frac{\lambda^4}{c}\left(\frac{c}{\kB T}\right)^l
  G_\mathrm{f}^{\alpha_1\alpha_2\cdots\alpha_l}
\end{equation}
be the dimensionless comoving extinction terms, the derivation in
appendix~\ref{app:extinction} gives us the following linear
transforms.
\begin{equation}
  \left[\begin{array}{l}
    \Gf\\\Gf^0\\\Gf^{00}\\
    \ \vdots\end{array}\right]
  = 8\pi
  \left[\begin{array}{cccc}
    X_2 & X_3 & X_4 & \cdots\\
    X_3 & X_4 & X_5 & \cdots\\
    X_4 & X_5 & X_6 & \cdots\\
    \vdots & \vdots & \vdots & \ddots
  \end{array}\right]
  \left[\begin{array}{l}{\bf}\\{\bf}_0\\{\bf}_{00}\\
    \ \vdots\end{array}\right],
\end{equation}
\begin{equation}
  \left[\begin{array}{l}
    \Gf^i\\\Gf^{0i}\\\ \vdots
  \end{array}\right]
  = \frac{8\pi}{3}
  \left[\begin{array}{ccc}
    X_4 & 2 X_5 & \cdots\\
    X_5 & 2 X_6 & \cdots\\
    \vdots & \vdots & \ddots
  \end{array}\right]
  \left[\begin{array}{l}{\bf}_i\\{\bf}_{0i}\\\ \vdots
  \end{array}\right],
\end{equation}
\begin{equation}
  \left[\begin{array}{l}\Gf^{ij}\\\ \vdots\end{array}\right]
  \!\!-\!\frac{1}{3}\!
  \left[\begin{array}{l}\Gf^{00}\\\ \vdots\end{array}\right]
  \!\!\dT^{ij}\! = \!\frac{16\pi}{15}\!\!
  \left[\begin{array}{cc}
    X_6 & \cdots\\
    \vdots & \ddots
  \end{array}\right]
  \left[\begin{array}{l}{\bf}_{ij}\\\ \vdots\end{array}\right].
\end{equation}
There is no freedom left in the above equations because $T$ and
$u^\alpha$ are both fixed as fluid properties.  We can apply inverse
Lorentz transform to boost the extinction terms back to the lab frame.

\section{Physical Implications in Closure Schemes}
\label{sec:physics}

It seems that we have all the equations to close the radiative moment
hierarchy.  Unfortunately, \citeauthor{Grad1949}'s moment method has
some arbitrariness in the choices of the fiducial frame velocity
$U^\alpha$ as we remarked in section~\ref{sec:flux}.  In order to get
some physical insights to constrain these arbitrariness, we need to
understand closure schemes at a more fundamental level.

\subsection{The Origin of Non-linearity}
\label{sec:non-linear}

Let us think deeper about the very successful ideal hydrodynamic
closure.  It only requites the left hand side of the Boltzmann
transport equation, which is linear, to derive the Euler equation.
Taking moments is also a linear operations.  So why is the resulting
equation non-linear?

Without loss of generality, we use non-relativistic hydrodynamics for
this discussion.  Let $m$ be the particle mass,
\begin{equation}
  v^i \equiv \frac{1}{m} \int d^3p f p^i
\end{equation}
be the fluid velocity, and $p'^i \equiv p^i - m v^i$ be the particle
momentum in the comoving frame.  The particle distribution $f$ is
always well approximated by a Maxwellian in the fluid comoving frame
because of collisions.  We can choose a linear closure in the comoving
frame, such as $p = (2/3) \rho e$, yet the non-linear inertial term
will always appear in the lab frame,
\begin{eqnarray}
  \frac{1}{m} \int d^3p f p^i p^j
  & = & m v^i v^j \int d^3p f + \frac{1}{m}\int d^3p f p'^i p'^j \nonumber\\
  & = & \rho v^i v^j + p g^{ij}.
\end{eqnarray}

Therefore, non-linearity does not just come from the closure
approximation.  It is a direct consequence of Galilean transforming
the second (or higher order) moment with a velocity that depends on
the distribution.  In other words, hydrodynamics is non-linear because
of \emph{Galilean symmetry} and \emph{fluid particle
  self-interaction}.

For radiation, we replace Galilean symmetry by \emph{Lorentz symmetry}
in the above reasoning.  Photons do not self-interact so they can only
be thermalized by external medium.  In the diffusion regime, a few
moments can fully describe the photon distribution function.  It does
not matter rather we use the fluid velocity $u^\alpha$ or some
radiative quantities to solve for the radiation fiducial velocity
$U^\alpha$.  Both choices can lead to fully relativistic and well
behave closures.

In the free-streaming regime, however, the radiating medium does not
contribute much.  On the one hand, setting $U^\alpha = u^\alpha$ is
non-sense.  On the other hand, employing the lab frame for linearity
causes closures, such as the $P_N$ closures, violates the Lorentz
symmetry.  We believe that Lorentz symmetry is an important property
\footnote{There is an exception.  If we need to implement a radiative
  transfer solver on a low resolution grid, the spatial discretization
  introduces fiducial frame dependence because of truncation error.
  It this case, it may not be a bad idea to just give up Lorentz
  invariance and use high order linear closures.  In some sense, this
  is the philosophy behind Lattice Boltzmann methods \citep[see, for
    example,][]{1993IJMPC...4..409S, 1997PhRvE..56.6811H,
    2000PhRvE..61.6546L}.} so we ensure that it is satisfied in the
proposed scheme.

\subsection{Unphysical Photon Self-Interaction}
\label{sec:interaction}

It is educational to study the simplest Lorentz invariant closure.  We
need at least one variable for the energy scale and three variables
for the fiducial velocity.  By symmetry, the radiative stress-energy
tensor must take the form
\begin{equation}
  R_\mathrm{r}^{\alpha\beta} = \left[\begin{array}{cccc}
  3 P & 0 & 0 & 0 \\
    0 & P & 0 & 0 \\
    0 & 0 & P & 0 \\
    0 & 0 & 0 & P
  \end{array}\right]
\end{equation}
in the fiducial frame.  Boosting it back
to the lab frame, we obtain the standard stress-energy tensor for
photon fluid,
\begin{equation}
  R^{\alpha\beta} = \frac{4}{c^2} P\,U^\alpha U^\beta + P g^{\alpha\beta}.
  \label{eq:radiative_fluid}
\end{equation}

Recalling equation~(\ref{eq:Rab}) and compare different components, we
obtain
\begin{eqnarray}
  E      & = & \left(\frac{4}{c^2}\,U^0 U^0 - 1\right) P, \\
  F^i    & = & \frac{4}{c}\,U^0 U^i P, \\
  P^{ij} & = & \left(\frac{4}{c^2}\,U^i U^j + g^{ij}\right) P.
\end{eqnarray}
Let us treat the radiative energy $E$ and flux $F^i$ as the
fundamental fields.  The closure problem reduces to solving $P^{ij}$
in terms of $E$ and $F^i$.  The solution is simply
\begin{equation}
  P^{ij} = \frac{F^i F^j}{c^2 (E + P)} + P g^{ij}
  \label{eq:rad_P_ij}
\end{equation}
with
\begin{equation}
  P = \frac{E}{3}\left(2\sqrt{1 - \frac{3 F^i F_i}{4 c^2 E^2}} - 1\right).
  \label{eq:rad_P}
\end{equation}

The above equations describe exactly the $M_1$ closure although the
derivation is different from the standard ones
\citep{1992A&A...265..345J, 2000A&A...356..559S, 2008PhRvD..78b4023F}.
In fact, our derivation is probably more general since the only
assumes are the number of fundamental fields, isotropy in the fiducial
frame, and Lorentz invariance in the closure.  There is no assumption
about the spectrum.

The above derivation is fully relativistic so the information
propagation speed is limited by the speed of light.  The closure
\emph{appears} to have correct diffusion and free-streaming limits.
However, the derivation implies that the photons always thermalize
themselves even without interacting with the radiating
medium\footnote{Recalling that the $M_1$ closure is originally
  developed to maximize the entropy \citep{1992A&A...265..345J}.}.  We
can interpret the truncation errors as \emph{unphysical photon
  self-interaction}.  Our closure problem now reduces to minimizing
this unphysical effect.

\subsection{Moment Decomposition}
\label{sec:decomposition}

There is no freedom left in the $M_1$ closure.  We have to introduce
more dynamic variables to reduce the unphysical photon
self-interaction.  The simplest trial is to include the zeroth moment
of the radiative transfer equation,
\begin{equation}
  {R^\alpha}_{;\alpha} = - G. \label{eq:rad0}
\end{equation}
The above equation and equation~(\ref{eq:rad1}) form a 5-field method.
Compare to $M_1$, the extra information governed by
equation~(\ref{eq:rad0}) allows us to go beyond the gray
approximation.  It is easy to derive the extinction terms by using
section~\ref{sec:extinction} or appendix~\ref{app:extinction}.  For
the purpose of this paper, we only focus at the closure problem.

The structure of the 5-field method is similar to relativistic
hydrodynamics~(\ref{eq:cont}) and (\ref{eq:Euler}), where the fluid
velocity is equal to the fiducial velocity.  I.e., the fiducial frame
is chosen so that the particle density flux vanishes.  The heat flux,
viscosity, etc are then defined in such a frame.  There is a special
name associated with this choice: the \citet{1940PhRv...58..919E}
decomposition.  Alternatively, \citet{1959flme.book.....L} proposed
shifting the fiducial velocity so that the heat flux vanishes in the
fiducial frame, $T_\mathrm{L}^{0i} = 0$, where the subscript
$_\mathrm{L}$ denotes the \citetalias{1959flme.book.....L} frame.

The fiducial velocity in the \citeauthor{1940PhRv...58..919E}
decomposition describes the particle flow, while in the
\citetalias{1959flme.book.....L} decomposition it describes the energy
flow.  When the radiation is not in local thermodynamic equilibrium,
the two decompositions result different fiducial frames and have
different level of unphysical photon self-interactions.  However, the
5-field method is too restrict to describe non-equilibrium effects.
The photon density flux $R^i$ is parallel to the photon energy flux
$R^{0i}$ in the fiducial frame,
\begin{equation}
  R_\mathrm{r}^i = \frac{W_3}{W_4} \frac{c}{\theta} R_\mathrm{r}^{0i}.
\end{equation}
The two decompositions are therefore identical.  We must use higher
order moment methods to reduce unphysical photon self-interaction.

\section{A Physically Motivated 14-Field Method}
\label{sec:14}

To describe non-equilibrium effects, we follow \citet{Grad1949} to
truncate the expansion~(\ref{eq:Grad}) at second order.  The photons
are described by an ultra-relativistic \emph{\citeauthor{Grad1949}'s
  distribution function},
\begin{equation}
  f^{(2)} \equiv \frac{2}{h^3}\, w(\xi)
  \Bigl( a + \xi a_\alpha n^\alpha +
  \xi^2 a_{\alpha\beta} n^\alpha n^\beta \Bigr) \approx f.
  \label{def:f2}
\end{equation}
Because $a_{\alpha\beta}$ is symmetric and has trace-free spatial
parts, there are only nine independent components.  Taking $a$ and
$a_\alpha$ into account, the polynomial has fourteen independent
coefficients.  They can be solved by the fourteen time dependent
fundamental fields $R^0$, $R^{0\alpha}$, and $R^{0\alpha\beta}$.

To derive our 14-field method, we first reduce the flux equations to
include only the non-vanishing coefficients,
\begin{eqnarray}
  \Rr^i & = & \frac{8\pi}{3}
    \Bigl(W_3 {\ar}_i + 2 W_4 {\ar}_{0i}\Bigr), \\
  \Rr^{ij} & = & \frac{16\pi}{15}
    \Bigl(W_5 {\ar}_{ij}\Bigr) + 
    \frac{1}{3} \Rr^{00} \dT^{ij}, \\
  \Rr^{ijk} & = & \frac{1}{5}
    \Bigl(\Rr^{00i}\dT^{jk} +
          \Rr^{00j}\dT^{ki} +
          \Rr^{00i}\dT^{jk}\Bigr).
          \label{eq:Rijk}
\end{eqnarray}
The dimensionless moments in the above equations are rescaled by the
energy scale $\theta$ chosen in equation~(\ref{eq:theta}).  Note that
$a$, $a_0$, and $a_{00}$ do not appear in the flux equations.
Therefore, we can skip equation~(\ref{eq:R0a0}) and just inverse the
reduced forms of equations~(\ref{eq:Riai}) and (\ref{eq:Rijaij}),
\begin{equation}
  \left[\begin{array}{l}
    \Rr^{0i} \\ \Rr^{00i}
  \end{array}\right] = \frac{8\pi}{3}
  \left[\begin{array}{cc}W_4 & 2 W_5 \\ W_5 & 2 W_6\end{array}\right]
  \left[\begin{array}{l}\ar^i \\ \ar^{0i}\end{array}\right],
\end{equation}
\begin{equation}
  \Rr^{0ij} - \frac{1}{3}\Rr^{000} \dT^{ij} 
  = \frac{16\pi}{15} W_6 {\ar}_{ij}.
\end{equation}
Eliminating \citeauthor{Grad1949}'s coefficients, we have
\begin{equation}
  \Rr^i =
  \frac{W_3 W_6 - W_4 W_5}{W_4 W_6 - W_5 W_5} \Rr^{0i} -
  \frac{W_3 W_5 - W_4 W_4}{W_4 W_6 - W_5 W_5} \Rr^{00i},
  \label{eq:Ri}
\end{equation}
\begin{equation}
  \Rr^{ij} =
  \frac{W_5}{W_6}\left(\Rr^{0ij} -
  \frac{1}{3}\Rr^{000} \dT^{ij}\right) +
  \frac{1}{3}\Rr^{00}\dT^{ij}.
  \label{eq:Rij}
\end{equation}
Equation~(\ref{eq:Ri}), (\ref{eq:Rij}), and (\ref{eq:Rijk}) almost
form a closure relation.  If we fixed the fiducial frame to the lab
frame, it would become a linear (but frame dependent) closure similar
to $P_2$.

Equation~(\ref{eq:Ri}) shows that the \citeauthor{1940PhRv...58..919E}
and \citetalias{1959flme.book.....L} decompositions result different
fiducial frames.  In addition, we can introduce a \emph{third order
  decomposition} so that the fiducial frame is chosen to satisfies the
requirement $\Rr^{00i} = 0$.  Equation~(\ref{eq:Ri}) becomes
\begin{eqnarray}
  \Rr^{0i} & = & \ \ \ 0.1035\Rr^{00i}
    \mbox{ in the \citeauthor{1940PhRv...58..919E} frame,}\\ 
  \Rr^{i}\ & = &      -0.0548\Rr^{00i}
    \mbox{ in the \citetalias{1959flme.book.....L} frame, and\ \ \ \ }
    \label{eq:LL}\\ 
  \Rr^{i}\ & = & \ \ \ 0.5299\Rr^{0i}\ \ 
    \mbox{ in the third order frame,}
\end{eqnarray}
respectively.  We can refer the \citeauthor{1940PhRv...58..919E} and
\citetalias{1959flme.book.....L} frames as first and second order
frames.

The sign difference between $\Rr^i$ and $\Rr^{00i}$ in
equation~(\ref{eq:LL}) has interesting meaning.  It indicates that the
\citetalias{1959flme.book.....L} fiducial velocity lies between the
other two.  Appendix~\ref{app:sensitivity} demonstrates that higher
order velocities are less sensitive to the distribution function.  The
non-linear terms have small contribution to the overall dynamics.
Therefore, we conjecture that higher order decompositions introduce
less unphysical photon-self interaction.  In order words, we propose
using the third order decomposition for the 14-field method.

\section{Conclusions}
\label{sec:conclusions}

In this paper, we propose a class of physically motivated closures for
radiative transfer and radiation hydrodynamics.  We start by showing
the advantages of frequency-integrated schemes, which are fully
relativistic.  The transport terms and extinction terms can be easily
evaluated in different reference frames.  We then apply
\citeauthor{Grad1949}'s moment method to compute the flux terms and
the extinction terms from the fundamental fields.  The truncated
\citeauthor{Grad1949}'s series is energy scale and fiducial frame
dependent.  For the extinction terms, the fluid comoving frame and
temperature are the natural choices.  For the flux terms, however, we
propose using high order decomposition to reduce unphysical photon
self-interaction as well as using photon number density to obtain the
energy scale.

We believe that this paper clarifies some implicit assumptions in
standard closures and points out the importance of fiducial frame.
Although we only present the 14-field method, it is straightforward to
derive arbitrarily high order closures from our formulas (see the
Appendixes).  With minimal modifications, our closures are also
applicable to neutrino transport and relativistic rarefied gas
dynamics.  We expect the methods derived from our new framework to
outperform existing ones.

Of course, there are many open questions associated with the proposed
schemes such as limiting behavior and linear stability of the theory.
We will address these issues in subsequent papers.  We are also
implementing the 14-field method and integrating it with some
hydrodynamic solvers.  We will perform verifications and validations
before applying the algorithms to study astrophysical systems.  As a
final remark, moment methods are not optimal for solving all problems
\citep{2006cmt..conf.....G}.  For example, for situations with strong
beaming, we are better off if we use other techniques such as ray
tracing \citep[see][and reference therein]{2002MNRAS.330L..53A,
  2007ApJ...671....1T, 2009MNRAS.393.1090F}.

\acknowledgements

This work is originally motivated by Feryal \"Ozel and Dimitrios
Psaltis.  It is a pleasure to thank Ramesh Narayan for his intuitions
on resolving many technical difficulties.  The author would also like
to thank Chris Fryer, Avi Loeb, George Rybicki, and Dimitri Mihalas
for helpful comments on moment methods.  The PiTP 2009 lectures by Jim
Stone and Mike Norman are extremely useful.  It is grateful to discuss
with Martin Pessah, Mark Dijkstra, Hy Trac, Mike Sekora, Sukanya
Chakrabarti, Robert Marcus, and Alexander Tchekhovskoy.  The author is
currently supported by an ITC fellowship.
\vspace{0pt} 

\appendix

\section{Grad's Expansion and Symmetric Trace-free Tensors}
\label{app:Grad}

Following the notations in section~\ref{sec:notations}, we use
$p^\alpha = p^0(1, n^i)$ to denote the four-momentum of a massless
particle, where $n^i$ is some spatial unit three-vector.  We assume
there exists energy scale $\theta$ to describe the particles in some
fiducial reference frame.  Note that $\theta$ is not necessary related
to the temperature.  Let $U^\alpha$ be the four-velocity of the
fiducial frame, we can define $\xi$ as a dimensionless measurement of
energy for each particle, $\xi = -p^\alpha U_\alpha/\theta$.  For
photon with frequency $\nu$, it reduces to $\xi = h\nu/\theta$ in the
fiducial frame, where $h$ is Planck's constant.

Considering a fixed position $x^\alpha$ in the fiducial frame, we can
expand the local particle distribution function in a weighted
multi-dimensional polynomial of $p^\alpha$,
\begin{equation}
  f = \frac{\gs}{h^3}\, w(\xi) \Bigl(\hat{a} +
  \hat{a}_{\beta_1 } p^{\beta_1} + \hat{a}_{\beta_1\beta_2 }
  p^{\beta_1} p^{\beta_2} + \hat{a}_{\beta_1\beta_2\beta_3}
  p^{\beta_1} p^{\beta_2} p^{\beta_3} + \cdots\Bigr).
  \label{eq:f}
\end{equation}
In the above equation, $\gs$ is the number of available states, which
equals 2 for ``photon gas''.  The function $w(\xi)$ is a dimensionless
\emph{weight}.  When it is well chosen, the integral for each term in
the polynomial is guaranteed to converge when we take moments.  Bear
in mind that $\xi$ is subject to Doppler shift.  Therefore, the above
expansion depends on the choice of the fiducial frame, i.e.,
$U^\alpha$, in addition to $\theta$.  The coefficient
$\hat{a}_{\beta_1\beta_2\cdots\beta_l}$ is a $l$-th rank tensor, which
has dimension $(c/\theta)^l$.  It is implicit in the notations that
these coefficients $\hat{a}$, $\hat{a}_{\beta_1}$,
$\hat{a}_{\beta_1\beta_2}$, ... depend on the position $x^\alpha$.

In the original \citeauthor{Grad1949}'s moment method for
non-relativistic and non-degenerated gas, because of the Maxwellian
distribution in equilibrium, the weighting function $w(\xi)$ is chosen
to be a multi-dimensional Gaussian, which turns out to be the weight
for the orthogonal conditions of Hermite polynomials, resulting a
Hermite series expansion.  By the same token, a natural weight for
relativistic non-degenerated gas is an exponentially decaying function
because of the Maxwell-J\"uttner distribution.  It corresponds to the
orthogonal condition of Laguerre polynomials.  Using this choice,
$\theta/\kB$ converges to temperature in the limit of local
thermodynamic equilibrium.  In this paper, however, we use $w(\xi) =
[\exp(\xi) - 1]^{-1}$ for radiation so the polynomial expansion
reduces to unity in the limit of black body radiation\footnote{The
  moment method proposed here is very general.  When applying to
  neutrino transport, we can chose $w(\xi) = [\exp(\xi) + 1]^{-1}$.
  This results $W_l = (1 - 2^{-l})\Gamma(l + 1)\zeta(l + 1)$ instead
  of equation~(\ref{def:Wl}).}.  It is not a weight for
classical/standard orthogonal polynomials.  Nevertheless, because of
its asymptotic behaviors, it is easy to verify that all the useful
integrals are guaranteed to converge.

From expansion~(\ref{eq:f}), it is clear that the tensor
$\hat{a}_{\beta_1\beta_2\cdots\beta_l}$ is totally symmetry.  In
addition, it has trace-free spatial parts because $p^\alpha$ has only
three independent components.  One way to verify this is to rewrite
the above expansion as
\begin{eqnarray}
  f & = & \frac{\gs}{h^3}\,w(\xi)\Bigl(a + 
    \xi   a_{\beta_1              } n^{\beta_1} + 
    \xi^2 a_{\beta_1\beta_2       } n^{\beta_1} n^{\beta_2} +
    \xi^3 a_{\beta_1\beta_2\beta_3} n^{\beta_1} n^{\beta_2} n^{\beta_3} + 
    \cdots\Bigr) \\
  & = & \frac{\gs}{h^3}\,w(\xi)\Bigl[
         \Bigl(a + \xi a_0 + \xi^2 a_{00} + \xi^3 a_{000} + \cdots\Bigr) + 
    \xi  \Bigl(a_{i_1} + 2 \xi a_{0 i_1} + 3 \xi^2 a_{0 0 i_1} + \cdots\Bigr)
      n^{i_1} +
    \nonumber\\ & & \ \ \ \ \ \ \ \ \ \ \ \ 
    \xi^2\Bigl(a_{i_1 i_2} + 3 \xi a_{0 i_1 i_2} + \cdots\Bigr)n^{i_1} n^{i_2} +
    \xi^3\Bigl(a_{i_1 i_2 i_3} + \cdots\Bigr) n^{i_1} n^{i_2} n^{i_3} + 
    \cdots\Bigr],
\end{eqnarray}
where the rescaled coefficients $a$, $a_{\beta_1}$,
$a_{\beta_1\beta_2}$, ... are all dimensionless.  We can now employ
the results in \citet{1980RvMP...52..299T} to conclude $a_{i_1 i_2
  \cdots i_l}$ must be a linear combination of symmetric trace-free
tensors.  Therefore, $\hat{a}_{\beta_1\beta_2\cdots\beta_l}$ has
trace-free spatial parts.  It is easy to see the close relationship
between (\ref{eq:f}) and hydrogen energy eigenstates by using the
identity between spherical harmonics and the basis set of symmetric
trace-free tensors.

Counting degrees of freedom is another way to verify the trace-free
property.  Recalling the zero moment of the ultra-relativistic
Boltzmann equation yields one dynamic equation; the first moments
yield four dynamic equations.  However, the second moments yield only
nine equations because photons are massless.  In order to match the
degrees of freedom at each order, $a_{\beta_1\beta_2\cdots\beta_l}$
has to be totally symmetric with trace-free spatial parts.

\section{Momentum Moments of Massless Particles}
\label{app:moments}

We are now ready to derive the equations for computing four-momentum
moments, $R^{\alpha_1\alpha_2\cdots\alpha_l}$, from an
ultra-relativistic one-particle distribution function $f$.  First, we
separate the $\xi$-dependence and angular-dependence of the particle
moment $p^\alpha$ by writing
\begin{eqnarray}
  R^{\alpha_1 \alpha_2 \cdots \alpha_l} & \equiv & 
    c \int\!\frac{d^3p}{p^0} f\,p^{\alpha_1} p^{\alpha_2} \cdots p^{\alpha_l}
  \nonumber\\ & = &
    c \int\!p^0dp^0\,d\Omega\, f\,p^{\alpha_1} p^{\alpha_2} \cdots p^{\alpha_l}
  \nonumber\\ & = &
    c \left(\frac{\theta}{c}\right)^{l+2} \int\!d\xi\,d\Omega\, f\, \xi^{l+1}
    n^{\alpha_1} n^{\alpha_2} \cdots n^{\alpha_l}
  \nonumber\\ & = & 
    \frac{\gs c}{h^3}
    \left(\frac{\theta}{c}\right)^{l+2} \int\!d\xi\,d\Omega\, w(\xi)
    \Bigl(a + \xi a_{\beta_1} n^{\beta_1} +
    \xi^2 a_{\beta_1\beta_2} n^{\beta_1} n^{\beta_2} + \cdots\Bigr)
    \xi^{l+1} n^{\alpha_1} n^{\alpha_2} \cdots n^{\alpha_l}.
    \label{eq:Ral}
\end{eqnarray}
We are free to rearrange the indices because
$R^{\alpha_1\alpha_2\cdots\alpha_l}$ is totally symmetric.  Let $m$ be
the total number of non-zero indices, without loss of generality, the
moment can be written in the form $R^{0 \cdots 0 i_1 i_2 \cdots i_m}$
by packing all the zeros to the beginning and leave the spatial
indices at the end.  We can separate all the $\xi$-dependence and
angular dependence,
\begin{eqnarray}
  R^{\alpha_1 \alpha_2 \cdots \alpha_l} 
    & = & R^{\scriptsize\overbrace{0 \cdots 0 i_1 i_2 \cdots i_m}^{
      \mbox{Totally $l$ indices}}}
  \nonumber\\ & = &
    \frac{\gs c}{h^3}
    \left(\frac{\theta}{c}\right)^{l+2} \int\!d\xi\,d\Omega\,
      w_{l+1} \Bigl(a + \xi a_{\beta_1} n^{\beta_1} +
         \xi^2 a_{\beta_1\beta_2} n^{\beta_1} n^{\beta_2} + \cdots
    \Bigr) n^{i_1} n^{i_2} \cdots n^{i_m}
  \nonumber\\ & = &
    \frac{\gs c}{h^3}
    \left(\frac{\theta}{c}\right)^{l+2} \int\!d\xi\,d\Omega\, \Bigl[ 
      w_{l+1} \Bigl(a + \xi a_0 + \xi^2 a_{00} + 
        \cdots \Bigr) n^{i_1} n^{i_2} \cdots n^{i_m}
        + \nonumber\\ & & \ \ \ 
      w_{l+2} \Bigl(a_{j_1} + 2\,\xi a_{0j_1} + 3\,\xi^3 a_{00j_1} + 
        \cdots \Bigr)
        n^{i_1} n^{i_2} \cdots n^{i_m} n^{j_1}
        + \nonumber\\ & & \ \ \ 
      w_{l+3} \Bigl(a_{j_1 j_2} + 3\,\xi a_{0j_1 j_2} +
        6\,\xi^2 a_{00j_1 j_2} + \cdots \Bigr)
        n^{i_1} n^{i_2} \cdots n^{i_m} n^{j_1} n^{j_2}
        + \nonumber\\ & & \ \ \ 
      w_{l+4} \Bigl(a_{j_1 j_2 j_3} + 4\,\xi a_{0j_1 j_2 j_3} +
        10\,\xi^2 a_{00j_1 j_2 j_3} + \cdots \Bigr)
        n^{i_1} n^{i_2} \cdots n^{i_m} n^{j_1} n^{j_2} n^{j_3} +
      \cdots
    \Bigr],
\end{eqnarray}
where we have introduced the shorthand $w_l \equiv w(\xi)\xi^l$.  It
turns out that there is a closed form solution for the $\xi$-integral,
\begin{equation}
  W_l \equiv \int_0^\infty d\xi w(\xi) \xi^l =
  \Gamma(l+1)\,\zeta(l+1) \mbox{ for photons,}
  \label{def:Wl}
\end{equation}
where $\Gamma$ is the Gamma function and $\zeta$ is the Riemann zeta
function.  Some numerical values are given here:
\begin{equation}
  W_1 \approx 1.645,\ \ \ W_2 \approx 2.404,\ \ \ W_3 \approx 6.494,\ \ \ 
  W_4 \approx 24.89,\ \ \ W_5 \approx 122.1,\ \ \ W_6 \approx 726.0,\ \ \
  \mbox{etc,\ \ \ for photons.}
  \label{eq:Wl_values}
\end{equation}
Therefore, with our shorthands,
\begin{eqnarray}
  R^{\scriptsize\overbrace{0 \cdots 0 i_1 i_2 \cdots i_m}^{
     \mbox{Totally $l$ indices}}} & = &
    \frac{\gs c}{h^3}
    \left(\frac{\theta}{c}\right)^{l+2} \int\!d\Omega\,\Bigl[
      \Bigl(C_0^0 W_{l+1} a + C_1^0 W_{l+2} a_0 + 
        C_2^0 W_{l+3} a_{00} + \cdots\Bigr)
        n^{i_1} n^{i_2} \cdots n^{i_m}
        + \nonumber\\ & & \ \ \ 
      \Bigl(C_1^1 W_{l+2} a_{j_1} + C_2^1 W_{l+3} a_{0 j_1} + 
        C_3^1  W_{l+4} a_{00 j_1} + \cdots\Bigr)
        n^{i_1} n^{i_2} \cdots n^{i_m} n^{j_1}
        + \nonumber\\ & & \ \ \ 
      \Bigl(C_2^2 W_{l+3} a_{j_1 j_2} + C_3^2 W_{l+4} a_{0 j_1 j_2} + 
        C_4^2  W_{l+5} a_{00 j_1 j_2} + \cdots\Bigr)
        n^{i_1} n^{i_2} \cdots n^{i_m} n^{j_1} n^{j_2}
        + \nonumber\\ & & \ \ \ 
      \Bigl(C_3^3 W_{l+4} a_{j_1 j_2 j_3} + C_4^3 W_{l+5} a_{0 j_1 j_2 j_3} + 
        C_5^3  W_{l+6} a_{00 j_1 j_2 j_3} + \cdots\Bigr)
        n^{i_1} n^{i_2} \cdots n^{i_m} n^{j_1} n^{j_2} n^{j_3} +
      \cdots
    \Bigr] \nonumber\\ & = &
    \frac{\gs c}{h^3}
    \left(\frac{\theta}{c}\right)^{l+2} \int\!d\Omega\,
      \sum_{p=0}^\infty \sum_{q=p}^\infty C_q^p W_{l + q + 1}
      a_{\scriptsize\underbrace{0 \cdots 0 j_1 j_2 \cdots j_p}_{
          \mbox{Totally $q$ indices}}}
      n^{i_1} n^{i_2} \cdots n^{i_m}
      n^{j_1} n^{j_2} \cdots n^{j_p}
    \nonumber\\ & = &
    \frac{\gs c}{h^3}
    \left(\frac{\theta}{c}\right)^{l+2}
      \sum_{p=0}^\infty \sum_{q=p}^\infty C_q^p W_{l + q + 1}
      a_{\scriptsize\underbrace{0 \cdots 0 j_1 j_2 \cdots j_p}_{
          \mbox{Totally $q$ indices}}}
      \int\!d\Omega\, 
      n^{i_1} n^{i_2} \cdots n^{i_m}
      n^{j_1} n^{j_2} \cdots n^{j_p}
  \label{eq:integral}
\end{eqnarray}
where $C_q^p$ denotes binomial coefficient $C_q^p \equiv q!/p!(q-p)!$.

The integrands in equation~(\ref{eq:integral}) are products of
different components of unit vectors.  We can employ
\citet{1980RvMP...52..299T}'s orthogonal condition to evaluate it:
\begin{equation}
  \frac{1}{4\pi} \int\!d\Omega\,n^{i_1} n^{i_2} \cdots n^{i_n} \equiv
  \frac{1}{n+1}\, \dT^{i_1 i_2 \cdots i_n},
\end{equation}
where we have defined $\dT^{i_1 i_2 \cdots i_n}$ be the Thorne delta
function of rank-$n$.  It carries interesting symmetric properties to
help us simplify the evacuation of the integral.  The function
vanishes when $n$ is odd.  For even $n$, it is defined to be the
completely symmetrized product of Kronecker delta functions.  For
completeness, we copy the formulas from \citet{1980RvMP...52..299T}:
\begin{equation}
  \dT^{i_1 i_2 \cdots i_n} =
  \delta^{(j_1 j_2}\cdots\delta^{j_{n-1} j_n)} \equiv
  \frac{1}{(n-1)!!} \sum_{k_2 k_4 \cdots k_n}
  \delta^{j_1 j_{k_2}} \delta^{j_{k_3} j_{k_4}} \cdots
  \delta^{j_{k_{n-1}} j_{k_n}}
\end{equation}
where
\begin{itemize}
  \item $k_2$ is summed from 2 to $n$;
  \item $k_3$ is the smallest integer not equal to 1 or $k_2$;
  \item $k_4$ is summed over all integers from 2 to $n$, not equal to
    $k_2$ or $k_3$;
  \item $k_5$ is the smallest integer not equal to 1 or $k_2$ or $k_3$
    or $k_4$;
  \item ...
\end{itemize}
We list all the non-zero Thorne deltas up to eight indices below:
\begin{eqnarray}
&&\dT = 1;\ \ \ 
  \dT^{ii} = 1;\ \ \ 
  \dT^{iiii} = 1,\ \ \
  \dT^{iijj} = \frac{1}{3};\ \ \
  \dT^{iiiiii} = 1,\ \ \
  \dT^{iiiijj} = \frac{1}{5},\ \ \
  \dT^{iijjkk} = \frac{1}{15};
\nonumber\\ && \ \ \ \ \ \ \ \ \ \ \ \ \ \ \ 
  \dT^{iiiiiiii} = 1,\ \ \
  \dT^{iiiiiijj} = \frac{1}{7},\ \ \
  \dT^{iiiijjjj} = \frac{3}{35},\ \ \
  \dT^{iiiijjkk} = \frac{1}{35}.
\end{eqnarray}
Note that the Einstein summation convention is not applied in the
above equations.  The symbols $i$, $j$, $k$ denotes indices that are
not equal to each other.  With the help of the Thorne delta function,
we obtain the expression for all moments
\begin{equation}
  R^{\alpha_1 \alpha_2 \cdots \alpha_l} \equiv
  R^{\scriptsize\overbrace{0 \cdots 0 i_1 i_2 \cdots i_m}^{
      \mbox{Totally $l$ indices}}}
  = 4\pi\,\frac{\gs c}{h^3}
    \left(\frac{\theta}{c}\right)^{l+2} 
    \sum_{p=0}^\infty \sum_{q=p}^\infty
    \frac{C_q^p W_{l + q + 1}}{m + p + 1}
    a_{\scriptsize\underbrace{0 \cdots 0 j_1 j_2 \cdots j_p}_{
        \mbox{Totally $q$ indices}}}
    \dT^{i_1 i_2 \cdots i_m j_1 j_2 \cdots j_p}.
  \label{eq:moment}
\end{equation}h
The number of zeros in the subscripts of $a_{0\cdots 0 j_1 j_2 \cdots
  j_p}$ increase as we sum over $q$.

The first few momentum moments have important physical meanings.  The
first moment $R^\alpha$ is the radiative four-flow.  Based on
equation~(\ref{eq:moment}),
\begin{equation}
  R^0 = \frac{8\pi c}{h^3} \left(\frac{\theta}{c}\right)^3 \Biggl[
    \Bigl( W_2 a + W_3 a_0 + W_4 a_{00} + \cdots\Bigr) +
    \frac{1}{3}\Bigl( W_4 a_{j_1j_2} + 3 W_5 a_{0j_1j_2} + 6 W_6
    a_{00j_1j_2} + \cdots\Bigr)\dT^{j_1 j_2} + \cdots \Biggr].
\end{equation}
Because $a_{\alpha_1\alpha_2\cdots\alpha_l}$ has trace-free spatial
part, the above equation reduces to
\begin{equation}
  R^0 = \frac{8\pi c}{\lambda^3}
    \Bigl( W_2 a + W_3 a_0 + W_4 a_{00} + \cdots\Bigr),
    \label{eq:R0}
\end{equation}
where we have defined the length scale $\lambda \equiv h c/\theta$,
which is proportional to the photon mean \emph{separation} (i.e.,
$n^{-1/3}$, it is a purely radiation properties and has nothing to do
with the material-dependent photon mean free path).  Similarly, the
spatial components of the radiative four-flow take the form
\begin{equation}
  R^i = \frac{8\pi c}{3\lambda^3}
    \Bigl( W_3 a_i + 2 W_4 a_{0i} + 3 W_5 a_{00i} + \cdots\Bigr).
\end{equation}
It becomes clear now, although a temperature is not well defined in
non-equilibrium relativistic statistic mechanics, the notion of mean
photon separation always exists.  It is possible to use this physical
interpretation to choose $\theta$.  The second moment
$R^{\alpha\beta}$ is the radiation stress-energy tensor.  Its
components take the form
\begin{eqnarray}
  R^{00} & = & \frac{8\pi c}{\lambda^3} \frac{\theta}{c}
    \Bigl( W_3 a + W_4 a_0 + W_5 a_{00} + \cdots\Bigr), \\
  R^{0i} & = & \frac{8\pi c}{3\lambda^3} \frac{\theta}{c}
    \Bigl( W_4 a_i + 2 W_5 a_{0i} + 3 W_6 a_{00i} + \cdots\Bigr), \\
  R^{ij} & = & \frac{16\pi c}{15\lambda^3} \frac{\theta}{c}
    \Bigl( W_5 a_{ij} + 3 W_6 a_{0ij} + 6 W_7 a_{00ij} + \cdots\Bigr)
    + \frac{1}{3} R^{00} \dT^{ij}.
\end{eqnarray}
For $R^{\alpha\beta\gamma}$,
\begin{eqnarray}
  R^{000} & = & \frac{8\pi c}{\lambda^3} \left(\frac{\theta}{c}\right)^2
    \Bigl( W_4 a + W_5 a_0 + W_6 a_{00} + \cdots\Bigr), \\
  R^{00i} & = & \frac{8\pi c}{3\lambda^3} \left(\frac{\theta}{c}\right)^2
    \Bigl( W_5 a_i + 2 W_6 a_{0i} + 3 W_7 a_{00i} + \cdots\Bigr), \\
  R^{0ij} & = & \frac{16\pi c}{15\lambda^3} \left(\frac{\theta}{c}\right)^2
    \Bigl( W_6 a_{ij} + 3 W_7 a_{0ij} + 6 W_8 a_{00ij} + \cdots\Bigr)
    + \frac{1}{3} R^{000} \dT^{ij}. \\
  R^{ijk} & = & \frac{16\pi c}{35\lambda^3} \left(\frac{\theta}{c}\right)^2
    \Bigl( W_7 a_{ijk} + 4 W_8 a_{0ijk} + 10 W_9 a_{00ijk} +
           \cdots\Bigr) +
  \frac{1}{5}\Bigl(R^{00i} \dT^{jk} +  R^{00j} \dT^{ki} + R^{00k} \dT^{ij}\Bigr)
\end{eqnarray}
Higher order moments can easily be obtained by using
equation~(\ref{eq:moment}).  Fixing the energy scale $\theta$ at each
point $x^\alpha$, the moments are simply linear transform of
\citeauthor{Grad1949}'s coefficients.  Hence, we can solve for the
coefficients by applying an inverse linear transfer.  Although the
zeroth moment $R$ is never used in radiative transfer, for
completeness, we give its expression here
\begin{equation}
  R \equiv c \int\!\frac{d^3p}{p^0} f
  = \frac{8\pi c^2}{\lambda^3 \theta}
    \Bigl( W_1 a + W_2 a_0 + W_3 a_{00} + \cdots\Bigr).
\end{equation}

\section{General Form of the Moment Extinction Terms}
\label{app:extinction}

The right hand side of the radiative transfer equation has the same
mathematical form as the linearized collision term of the
Bhatnagar-Gross-Krook model \citep[see standard textbook such
  as][]{1987stme.book.....H, 2002rbet.book.....C}.  We start deriving
its moments by defining
\begin{equation}
  G^{\alpha_1 \alpha_2 \cdots \alpha_l} \equiv
  c \int\!\frac{d^3p}{p^0}\,
    \emn{x} (f - s) p^{\alpha_1} p^{\alpha_2} \cdots p^{\alpha_l} =
  c \left(\frac{\kB T}{c}\right)^{l + 2} \int\!d\xi\,d\Omega\,
  \emn{x} (f - s) \xi^{l + 1} n^{\alpha_1} n^{\alpha_2} \cdots n^{\alpha_l},
\end{equation}
where $s$ denotes the source distribution function and $\emn{x}$ is
the Lorentz invariant extension coefficient.  Similar to the
appendix~\ref{app:moments}, we define $\xi \equiv h\nu/\kB T$ and
replace the integral over $\nu$ by the integral over $\xi$.  We assume
thermal radiation so
\begin{equation}
  s \equiv \frac{\emn{e}}{\emn{x}} =
  \frac{\gs/h^3}{e^{p^\alpha u_\alpha/\kB T} - 1} =
  \frac{\gs}{h^3}\,w(\xi),
\end{equation}
where we further replace $U^\alpha$ by the fluid four-velocity
$u^\alpha$ and $\theta$ by the fluid temperature $\kB T$ for $w(\xi)$.
Substituting \citeauthor{Grad1949}'s series for $f$, the extinction
terms become
\begin{equation}
  G^{\alpha_1 \alpha_2 \cdots \alpha_l} \equiv
  \frac{\gs c}{h^3}
  \left(\frac{\kB T}{c}\right)^{l + 2} \int\!d\xi\,d\Omega\,
  \emn{x} w_{l+1} \Bigl(a - 1 + \xi a_{\beta_1} n^{\beta_1} +
  \xi^2 a_{\beta_1\beta_2} n^{\beta_1} n^{\beta_2} + \cdots
  \Bigr) n^{\alpha_1} n^{\alpha_2} \cdots n^{\alpha_l}.
\end{equation}

The zeroth and first moments of the collision term have clear physical
meanings.  When $l = 0$, $G$ is the photon extinction rate; while for
$l = 1$, $G^\alpha$ is the radiative four-force.  Higher order moment
such as $G^{\alpha\beta}$ are simply called the extinction term of the
third moment because of the balance equations.  We evaluate
$G^{\alpha_1 \alpha_2 \cdots \alpha_l}$ in the fluid comoving frame so
that $u^\alpha$ has only the temporal component.

The Lorentz invariant extension coefficient $\emn{x} \equiv (h\nu/c)
\chi_\nu$ depends on the radiative process.  Consider free-free,
bound-free, or electron scatter, we notice that each of them is
proportional to some power of temperature and frequency, $\chi_\nu
\propto T^\phi\nu^\psi$, in standard approximations.  Therefore, we
assume for the general form,
\begin{equation}
  \chi_\nu = \frac{\tau(\xi)}{\lambda}
    \left(\frac{\kB T}{\me c^2}\right)^{\phi + \psi - 1},
\end{equation}
where $\lambda \equiv h c / \kB T$ and $\tau(\xi)$ is some
dimensionless function in $\xi$.  The momentum moment of the collision
term due to a particular radiative process becomes
\begin{equation}
  G^{\alpha_1 \alpha_2 \cdots \alpha_l} = 
  \frac{\gs c}{\lambda^4}
  \left(\frac{\kB T}{c}\right)^{l}
  \left(\frac{\kB T}{\me c^2}\right)^{\phi + \psi - 1}
  \int\!d\xi\,d\Omega\, \tau(\xi) w_{l+2}
  \Bigl(a - 1 + \xi a_{\beta_1} n^{\beta_1} +
  \xi^2 a_{\beta_1\beta_2} n^{\beta_1} n^{\beta_2} + \cdots\Bigr)
  n^{\alpha_1} n^{\alpha_2} \cdots n^{\alpha_l}.
  \label{eq:Pal}
\end{equation}
The expression is very similar to equation~(\ref{eq:Ral}).  We can
further define
\begin{equation}
  b \equiv a - 1 \mbox{,\ \ \ }
  b_{\beta_1} \equiv a_{\beta_1} \mbox{,\ \ \ }
  b_{\beta_1\beta_2} \equiv a_{\beta_1\beta_2} \mbox{,\ \ \ ...}
\end{equation}
and the shorthand
\begin{equation}
  X_l \equiv \int_0^\infty d\xi\,\tau(\xi)\,w(\xi)\,\xi^l.
  \label{def:Xl}
\end{equation}
Note that $l$ is not necessary an integer in the above definition.
Fortunately, the values of $X_l$ are well define as soon as $\chi_\nu$
does not grow exponentially.  Follow the same procedures in
appendix~\ref{app:moments}, we can easily deduce the following
expression
\begin{equation}
  G^{\alpha_1 \alpha_2 \cdots \alpha_l} \equiv
  G^{\scriptsize\overbrace{0 \cdots 0 i_1 i_2 \cdots i_m}^{
      \mbox{Totally $l$ indices}}}
  = 4\pi\,\frac{\gs c}{\lambda^4}
  \left(\frac{\kB T}{c}\right)^{l}
  \left(\frac{\kB T}{\me c^2}\right)^{\phi + \psi - 1}
  \sum_{p=0}^\infty \sum_{q=p}^\infty
  \frac{C_q^p X_{l + q + 2}}{m + p + 1}
  b_{\scriptsize\underbrace{0 \cdots 0 j_1 j_2 \cdots j_p}_{
      \mbox{Totally $n$ indices}}}
  \dT^{i_1 i_2 \cdots i_m j_1 j_2 \cdots j_p}.
  \label{eq:extinction}
\end{equation}

The zeroth order moment is the photon extinction rate.  It is simply
\begin{equation}
  G = \frac{8\pi c}{\lambda^4} 
  \left(\frac{\kB T}{\me c^2}\right)^{\phi + \psi - 1}
  \Bigl( X_2 b + X_3 b_0 + X_4 b_{00} + \cdots\Bigr).
  \label{eq:P}
\end{equation}
The radiative four-force takes the form
\begin{eqnarray}
  G^{0} & = &
  \frac{8\pi c}{\lambda^4} \frac{\kB T}{c} 
    \left(\frac{\kB T}{\me c^2}\right)^{\phi + \psi - 1}
    \Bigl( X_3 b + X_4 b_0 + X_5 b_{00} + \cdots\Bigr), \\
  G^{i} & = &
  \frac{8\pi c}{3\lambda^4} \frac{\kB T}{c} 
    \left(\frac{\kB T}{\me c^2}\right)^{\phi + \psi - 1}
    \Bigl(X_4 b_i + 2 X_5 b_{0i} + 3 X_6 b_{00i} + \cdots\Bigr).
\end{eqnarray}
For the second moment $G^{\alpha\beta}$, we have
\begin{eqnarray}
  G^{00} & = &
  \frac{8\pi c}{\lambda^4} \left(\frac{\kB T}{c}\right)^2
  \left(\frac{\kB T}{\me c^2}\right)^{\phi + \psi - 1}
  \Bigl(X_4 b + X_5 b_0 + X_6 b_{00} + \cdots\Bigr), \\
  G^{0i} & = &
  \frac{8\pi c}{3\lambda^4} \left(\frac{\kB T}{c}\right)^2
  \left(\frac{\kB T}{\me c^2}\right)^{\phi + \psi - 1}
  \Bigl(X_5 b_i + 2 X_6 b_{0i} + 3 X_7 b_{00i} + \cdots\Bigr),\\
  G^{ij} & = & 
  \frac{16\pi c}{15\lambda^4} \left(\frac{\kB T}{c}\right)^2
  \left(\frac{\kB T}{\me c^2}\right)^{\phi + \psi - 1}
  \Bigl(X_6 b_{ij} + 3 X_7 b_{0ij} + 6 X_8 b_{00ij}
  + \cdots\Bigr) + \frac{1}{3} G^{00} \dT^{ij}.
\end{eqnarray}
Other higher order moments can be obtained by using
equation~(\ref{eq:extinction}).  If there are more than one radiative
process in the problem, we can compute the extinction terms for each
process and then sum over the results.

\section{Sensitivity of Fiducial Velocity}
\label{app:sensitivity}

We consider a very simple radiative transfer problem.  Assuming there
are only two streams of photons moving along $\mathbf{\hat{x}}^1$-axis
in opposite directions, the photon distribution function is
\begin{equation}
  f = \frac{\gs}{h^3} \Bigl[
    n_+ \delta(\nu - \nu_+) \delta^3(n^i - \delta^{i1}) +
    n_- \delta(\nu - \nu_-) \delta^3(n^i + \delta^{i1})
  \Bigr],
\end{equation}
where $n_\pm$ and $\nu_\pm$ are the density and frequency for the two
streams, $\delta(\nu)$ and $\delta^3(n^i)$ are the one- and
three-dimensional Dirac delta functions.  It is trivial to evaluate
its momentum moments.  The non-vanishing terms are
\begin{equation}
  R^{\scriptsize\overbrace{0\cdots00}^{\mbox{$l$ indices}}} \propto
  \left(n_+\nu_+^{l+1} + n_-\nu_-^{l+1}\right) \mbox{\ ,\ \ \ }
  R^{\scriptsize\overbrace{0\cdots01}^{\mbox{$l$ indices}}} \propto
  \left(n_+\nu_+^{l+1} - n_-\nu_-^{l+1}\right) \mbox{\ ,\ \dots}
\end{equation}
For an $l$-th order decomposition, we choose the fiducial reference
frame so that the $l$-th order flux $R^{0\cdots01}$ vanish.  This is
equivalent to solve for $\beta \equiv U^1 / c$ in the equation
\begin{equation}
  n_+ \left(\ \nu_+ \sqrt{\frac{1 - \beta}{1 + \beta}}\ \ \right)^{l+1} = 
  n_- \left(\ \nu_- \sqrt{\frac{1 + \beta}{1 - \beta}}\ \ \right)^{l+1}.
\end{equation}

Suppose $\beta_l$ is a solution to the above equation.  To see how the
fiducial frame depends on the distribution function, we suppose there
is a small change to the photon number density $n_\pm$.  The
corresponding change in the fiducial velocity is related to the
derivative
\begin{equation}
  \left.\frac{d\beta}{dn_\pm}\right|_\mathrm{\beta = \beta_l} =
  \pm \frac{1-\beta_l^2}{2 n_\pm (1+l)}.
\end{equation}
In the limit $l \rightarrow \infty$, the derivative $d\beta/dn_\pm
\rightarrow 0$, which recovers the linear closure.  Although the
general situation is more complicated, it is sensible to conjecture
that the fiducial velocity obtained from higher order decompositions
are less sensitive to the distribution function.  Hence, the
non-linear Lorentz transforms should introduce less unphysical photon
self-interaction.

\bibliography{ms,my}

\end{document}